\documentclass[aps,prl,reprint,superscriptaddress,nofootinbib,longbibliography]{revtex4-2}
\usepackage{amsmath}
\usepackage{amssymb}
\usepackage{amsthm}
\usepackage[figuresright]{rotating}
\usepackage[colorlinks=true,linkcolor=blue,anchorcolor=red,citecolor=blue, urlcolor=blue]{hyperref}
\usepackage{bm}
\usepackage{graphicx}
\usepackage{color}
\usepackage{multirow}
\usepackage{extarrows}
\usepackage{array}
\usepackage{appendix}
\usepackage{footnote}

\usepackage{cases}
\usepackage[normalem]{ulem}
\begin{document}

\title{Scaling analysis of quantum geometry in second-order nonlinear transport}

\author{Zhen-Hao Gong$^\dagger$}
\affiliation{State Key Laboratory of Quantum Functional Materials, Department of Physics, Guangdong Provincial Key Laboratory of Topological Matter, and Guangdong Basic Research Center of Excellence for Quantum Science, Southern University of Science and Technology (SUSTech), Shenzhen 518055, China}

\author{Z. Z. Du$^\dagger$}
\affiliation{Quantum Science Center of Guangdong-Hong Kong-Macao Greater Bay Area (Guangdong), Shenzhen 518045, China}

\author{Hai-Peng Sun}
\affiliation{Shenzhen Key Laboratory of Ultraintense Laser and Advanced Material Technology, Center for Intense Laser Application Technology, and College of Engineering Physics, Shenzhen Technology University, Shenzhen 518118, China}

\author{Hai-Zhou Lu}
\email[Corresponding author: ]{luhz@sustech.edu.cn}
\affiliation{State Key Laboratory of Quantum Functional Materials, Department of Physics, Guangdong Provincial Key Laboratory of Topological Matter, and Guangdong Basic Research Center of Excellence for Quantum Science, Southern University of Science and Technology (SUSTech), Shenzhen 518055, China}
\affiliation{Quantum Science Center of Guangdong-Hong Kong-Macao Greater Bay Area (Guangdong), Shenzhen 518045, China}

\author{X. C. Xie}
\affiliation{International Center for Quantum Materials, School of Physics, Peking University, Beijing 100871, China}
\affiliation{Interdisciplinary Center for Theoretical Physics and Information Sciences (ICTPIS), Fudan University, Shanghai 200433, China}
\affiliation{Hefei National Laboratory, Hefei 230088, China}

\begin{abstract}
Quantum geometry encodes the structure of the Hilbert space of Bloch states and can be accessed through nonlinear transport. Yet, disorder-induced mechanisms generically contribute to nonlinear transport, making it difficult to isolate quantum-geometric contributions in experiments. Here we systematically enumerate geometric and disorder-induced mechanisms of the second-order nonlinear Hall effect and derive a scaling law that expresses the nonlinear Hall conductivity as a polynomial of the linear longitudinal conductivity. Crucially, each mechanism carries a distinct “weight fingerprint” in the polynomial, enabling a quantitative disentanglement of quantum geometry from disorder backgrounds in existing experiments, both with and without time-reversal symmetry. Our results provide an implementable workflow for identifying quantum-geometric contributions in nonlinear-transport measurements.
\end{abstract}

\maketitle
\begingroup
\renewcommand{\thefootnote}{\ensuremath{\dagger}}
\footnotetext{Zhen-Hao Gong and Z. Z. Du contributed equally to this work.}
\endgroup
\begingroup
\renewcommand{\thefootnote}{*}
\footnotetext{Corresponding author: \href{mailto:luhz@sustech.edu.cn}{luhz@sustech.edu.cn}}
\endgroup

\section{Introduction}
\label{Sec:Intro}

\begin{figure*}[htbp]
\centering
\includegraphics[width=0.95\textwidth]{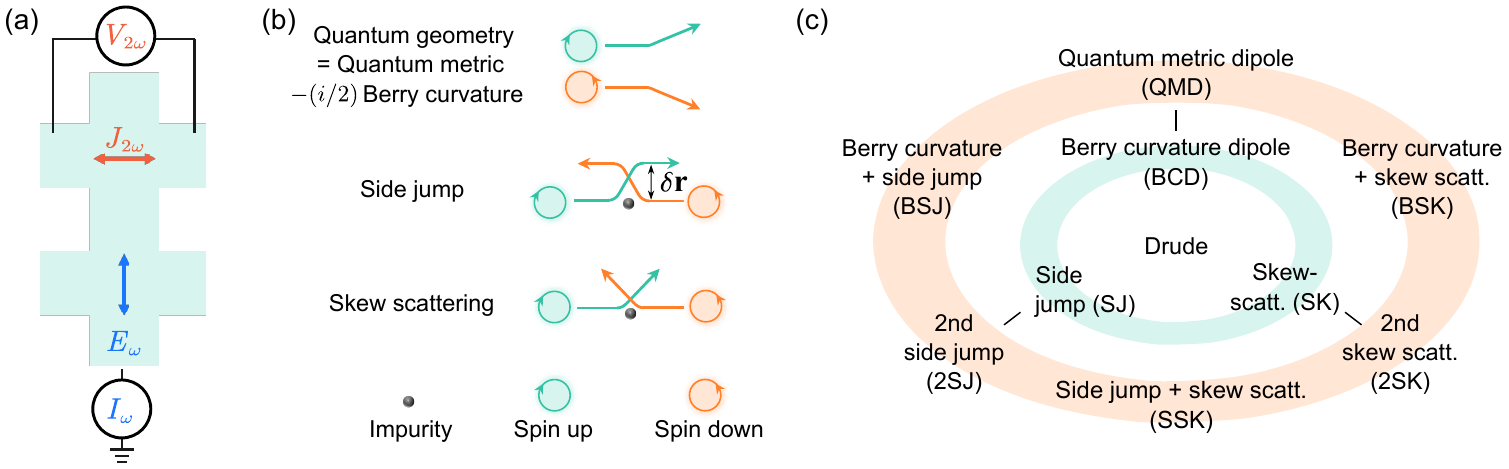}
\setlength{\abovecaptionskip}{-1pt}
\caption{Mechanisms of the second-order nonlinear transport. (a) Nonlinear Hall effect is measured experimentally as a double-frequency transverse voltage $V_{2\omega}$ driven by a low-frequency current $I_\omega$ (10–1000 Hz), or calculated theoretically as a current density $J_{2\omega}$ driven by an electric field $E_\omega$. (b) The building blocks for the nonlinear transport, including quantum geometry (= quantum metric $- i/2$ Berry curvature), SJ (which induces a lateral coordinate shift of spin-up and spin-down electrons oppositely by $\delta\mathbf{r}$), and SK (which asymmetrically deflects spin-up and spin-down electrons). (c) Hierarchy of the second-order nonlinear transport. (i) The Berry curvature dipole (BCD) \cite{Fu2015PRL,Low15prb}, SJ, and SK mechanisms \cite{Du2019NC,Du2021NC} survive time-reversal ($\mathcal{T}$) symmetry, require broken spatial inversion ($\mathcal{P}$) symmetry, and are forbidden by $\mathcal{PT}$ symmetry. (ii) The Drude, quantum metric dipole (QMD) \cite{GaoY14prl,Xiao2021PRL,Yang2021PRL} mechanisms as well as second-order skew scattering (2SK), second-order side jump (2SJ), side-jump-plus-skew-scattering (SSK), Berry-curvature-plus-side-jump (BSJ), and Berry-curvature-plus-skew scattering (BSK), part of which has been known as the anomalous skew scattering~\cite{Song2023PRL}, require broken $\mathcal{T}$ and $\mathcal{P}$ symmetries but survive the combined $\mathcal{PT}$ symmetry.}
\label{Fig:Mechanisms}
\end{figure*}

The nonlinear Hall effect manifests as a transverse voltage nonlinearly dependent on the longitudinal driving current [Fig.~\ref{Fig:Mechanisms}(a)] \cite{GaoY14prl,Fu2015PRL,Low15prb,Ma2019Nature,Du2018PRL,Kang2019NatMat,Du2021NRP,Xiao2021PRL,Yang2021PRL,Xu2023Science,Yan2023Nature}.
It has attracted broad interest due to its potential as a spectroscopic tool \cite{Du2019NC,Xiao2019PRB,Du2021NC,Tiwari2021NC,Duan2022PRL,sinha22np,Kaplan23nc,Mao2023NC,Jin23prl,Yokouchi23prl,Huang23prl,Fang24prl,Su24prl,Kaplan24prl,ChenR24prbl,HeP24nn,Wang24prl,makushko24ne,Lihm24prl} and in device applications such as wireless rectification \cite{Fu2020SA,Kumar2021NN}, as well as its connections to nonlinear optics \cite{Song2023PRL,Ghosh24sc,Shin24prl,Chang25prl}.
Notably, the second-order response can originate from the quantum metric \cite{GaoY14prl, Xiao2021PRL,Yang2021PRL} and Berry curvature \cite{Fu2015PRL,Low15prb,Du2018PRL}, which correspond to the real and imaginary parts of quantum geometry, respectively \cite{Provost80cmp,Resta11epjb,parameswaran2012, roy2014,peotta2015, Torma22nrp,Torma23prl,Balazs23pra,HuJM24arXiv}. Quantum geometry quantifies distances between quantum states and can provide geometric insights into challenges in condensed matter \cite{Liu24nsr}, in particular in the fractional Chern insulator \cite{parameswaran2012, roy2014, peotta2015} and flat-band superconductivity \cite{Torma22nrp,Torma23prl}.
However, disorder-induced mechanisms, such as side jump (SJ) that shifts coordinates of electrons sideways and skew scattering (SK) that asymmetrically deflects electrons [Fig.~\ref{Fig:Mechanisms}(b)] \cite{Nagaosa2010RMP}, are often comparable and must be accounted for in the nonlinear transport \cite{Du2019NC,Fu2020SA,Song2023PRL,Atencia23prbl,Huang25prb}.
Therefore, a central open issue is how to unambiguously identify quantum geometry from the disorder-induced mechanisms, despite a huge amount of experimental data.

In this work, we derive the formulas for all possible mechanisms of second-order nonlinear transport by treating geometry and disorder on an equal footing. The formulas allow us to derive a scaling law that expresses the nonlinear Hall conductivity as a polynomial function of the linear longitudinal conductivity. More importantly, we find each mechanism has a distinct weight distribution of the polynomial terms, which allows us to identify quantum geometry from disorder-induced mechanisms in the recent experiments with and without $\mathcal{T}$ symmetry. The formulas also allow us to determine nonzero nonlinear conductivity elements for materials categorized by the magnetic point groups.  Our theory is not only a comprehensive description of the second-order nonlinear transport, but also provides a quantitative tool and implementable workflow for
identifying quantum-geometry–driven nonlinear responses in future experiments and applications.

\section{Mechanisms of second-order nonlinear transport}

A Hall effect occurs when electrons driven by an electric field are deflected to the transverse direction. In the absence of an external magnetic field, three mechanisms may lead to the Hall effect linear in electric field $\vec{E}$: Berry curvature, SJ, and SK \cite{Nagaosa2010RMP}, as shown in Fig.~\ref{Fig:Mechanisms}(b).
The Berry curvature \cite{Xiao2010RMP} describes the curvature of the Hilbert space of the Bloch states, which thereby prevents electrons from moving straight in a perfect crystal. Moreover, the Berry-curvature contribution itself acquires an electric-field-induced geometric correction, governed by the quantum metric \cite{GaoY14prl,Xiao2021PRL,Yang2021PRL}.
In the presence of disorder, that is, imperfection of the crystal, SJ shifts the coordinates of spin-up and spin-down electrons sideways in opposite directions,
and SK asymmetrically deflects spin-up and spin-down
electrons due to the spin-orbit coupling. 
These mechanisms and mixing of them can capture all contributions up to second order in the electric field (i.e., $\propto E^2$) within the semiclassical framework. The hierarchy of all mechanisms for the nonlinear Hall effect is categorized in Fig.~\ref{Fig:Mechanisms}(c) according to three different symmetries, i.e., the $\mathcal{T}$, $\mathcal{P}$, and combined $\mathcal{PT}$ symmetries.

Experimentally, the nonlinear Hall effect is measured as a double-frequency transverse voltage in response to a low-frequency current (10–1000 Hz). Theoretically, the nonlinear conductivity $\chi_{abc}$ is defined in terms of the
double-frequency current density
\begin{eqnarray}
J_a(2\omega)=\chi_{a b c} E_b(\omega)
E_c(\omega),
\end{eqnarray}
in response to low-frequency electric fields $E_b(\omega)$ and $E_c(\omega)$, where $\omega$ is the frequency, $a,b,c\in \{x,y,z\}$. $\chi_{abc}$ can be calculated using the semiclassical theory to treat quantum geometry (including the Berry curvature and quantum metric) and disorder on an equal footing. We find that the second-order nonlinear conductivity $\chi_{abc}$ can be decomposed as
\begin{eqnarray}\label{Eq:chi_abc}
\chi_{abc}
&=&\chi^\mathrm{Drude}\nonumber\\
&&+\chi^{\mathrm{BCD}}
+\chi^{\mathrm{SJ}}
+\chi^{\mathrm{SK}}\nonumber\\
&&+ \chi^{\mathrm{QMD}}+\chi^{\mathrm{2SK}}+\chi^{\mathrm{2SJ}}+\chi^{\mathrm{SSK}}
+\chi^{\mathrm{BSJ}}+\chi^{\mathrm{BSK}},\nonumber\\
\end{eqnarray}
where $\chi^{\mathrm{2SK}}, \chi^{\mathrm{2SJ}}
, \chi^{\mathrm{SSK}}
, \chi^{\mathrm{BSJ}}$, and part of $\chi^{\mathrm{BSK}}$ are derived in this work, complementing five previously known mechanisms \cite{GaoY14prl,Fu2015PRL,Low15prb,Du2019NC,Fu2020SA,Xiao2021PRL,Yang2021PRL,Song2023PRL,Huang25prb}. 
Among the mechanisms, QMD and BCD are particularly notable as they directly probe quantum geometry that measures the distance between quantum states.
All terms on the first line of Eq.~(\ref{Eq:chi_abc}) have the same symmetry under which the quantum metric emerges in the nonlinear transport. The derivation, explicit formulas, and symmetries of the mechanisms can be found in \hyperlink{Appendix}{Appendix}.

\begin{table}[bthp]
\caption{Weights of the polynomial terms in the scaling law Eq.~(\ref{eq:scaling-dynamic}) for each of the mechanisms of the second-order nonlinear Hall effect in Fig.~\ref{Fig:Mechanisms}. The mechanisms without (with) an asterisk (do not) require broken $\mathcal{T}$ symmetry.}
\label{Tab:scaling}
\centering
\small
\setlength{\tabcolsep}{0.13cm}
\begin{tabular}{cccccccccccccc}
\hline
\textbf{Mechanism}&$\mathcal{C}_4\sigma_{xx}^4$&:&$\mathcal{C}_3\sigma_{xx}^3$&:&$\mathcal{C}_2\sigma_{xx}^2$&:&$\mathcal{C}_1\sigma_{xx}$&:&$\mathcal{C}_0$  
\\
\hline 
Drude  &  $\phantom{-}0$&:&$\phantom{-}0$&:&$\phantom{-}1$&:&$\phantom{-}0$&:&$\phantom{-}0$    
\\
BCD* &  $\phantom{-}0$&:&$\phantom{-}0$&:&$\phantom{-}0$&:&$\phantom{-}1$&:&$\phantom{-}0$    
\\
SJ* &  $\phantom{-}0$&:&$\phantom{-}0$&:&$\phantom{-}1$&:&$-1$&:&$\phantom{-}0$     
\\
SK* & $\phantom{-}0$&:&$\phantom{-}1$&:&$-2$&:&$\phantom{-}1$&:&$\phantom{-}0$    
\\
QMD  &  $\phantom{-}0$&:&$\phantom{-}0$&:&$\phantom{-}0$&:&$\phantom{-}0$&:&$\phantom{-}1$    
\\
BSJ &   $\phantom{-}0$&:&$\phantom{-}0$&:&$\phantom{-}0$&:&$\phantom{-}1$&:&$-1$    
\\
2SK &   $\phantom{-}1$&:&$-4$&:&$\phantom{-}6$&:&$-4$&:&$\phantom{-}1$     
\\
SSK &   $\phantom{-}0$&:&$\phantom{-}1$&:&$-3$&:&$\phantom{-}3$&:&$-1$     
\\
\hline
BSK, 2SJ  &  $\phantom{-}0$&:&$\phantom{-}0$&:&$\phantom{-}1$&:&$-2$&:&$\phantom{-}1$    \\
\hline
\end{tabular}
\end{table}

\section{Scaling law of second-order nonlinear transport}

Because disorder-induced mechanisms are present, quantitatively identifying the quantum-geometry contributions to nonlinear transport remains challenging. Scaling laws have proven successful in distinguishing mechanisms in linear Hall effects \cite{Nagaosa2010RMP,Tian2009PRL,Hou2015PRL}.
We derive a scaling law for the second-order nonlinear transport as a polynomial relation between the transverse nonlinear Hall conductivity $\chi_{yxx}$ and longitudinal linear conductivity $\sigma_{xx}$
\begin{eqnarray}\label{eq:scaling-dynamic}
\chi_{yxx}
=\mathcal{C}_4 \sigma_{xx}^4+\mathcal{C}_3 \sigma_{xx}^3+\mathcal{C}_2 \sigma_{xx}^2+\mathcal{C}_1 \sigma_{xx}+\mathcal{C}_0,
\end{eqnarray}
where in experiments 
$\chi_{yxx}= 
V^{2\omega}_{y}\sigma_{xx}L^2/(V^{\omega}_{x})^{2}W$, $L$ is the length, $W$ is the width in two-dimensional (2D) systems or cross-sectional area in three-dimensional (3D) systems, $V^{2\omega}_{y}$ is the transverse double-frequency voltage, and $V^{\omega}_{x}$ is the longitudinal single-frequency voltage. The scaling law is also found for $\mathcal{T}$-odd nonlinear Hall effect \cite{Huang25prb}. In this work, we further find distinct weight distributions of the $\mathcal{C}_i\sigma_{xx}^i$ terms for all the mechanisms, as shown in Table \ref{Tab:scaling}, 
which could serve as signatures to distinguish the mechanisms in experiments. Except that the BSK and 2SJ mechanisms have the same weight distribution, the remaining 8 out of 10 mechanisms have unique weight distributions, such as QMD or 2SK. The workflow of identifying the mechanisms is as follows. (i) Fit $\chi_{yxx}$ as a polynomial of $\sigma_{xx}$ using Eq. (3). (ii) Identify the underlying mechanisms by matching the fitted weights to Table \ref{Tab:scaling}.

\begin{figure*}[htpb]
\centering
\includegraphics[width=0.95\textwidth]
{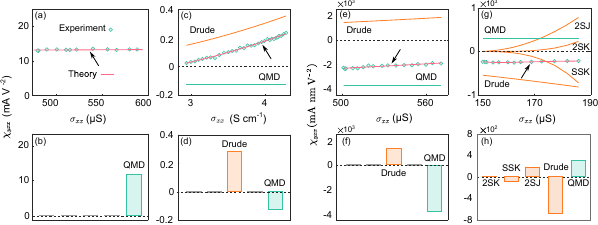}
   \caption{Scaling analysis of the mechanisms for the nonlinear Hall effect in MnBi$_2$Te$_4$ thin films. (a) Fitted nonlinear Hall conductivity $\chi_{yxx}$ (red solid curve) as a function of the linear longitudinal conductivity $\sigma_{xx}$, for the experimental data (circles) of a 6-septuple-layer MnBi$_2$Te$_4$ thin film strained by black phosphorus in the temperature range $T=2$–$13$ K \cite{Xu2023Science}.  (b) Weights of the five $\mathcal{C}_i\sigma_{xx}^i$ terms in Eq.~(\ref{eq:scaling-dynamic}) for the data from (a) at $\sigma_{xx}=540\ \mu $S (indicated by the arrow).
[(c) and (d)] For a 6-septuple-layer MnBi$_2$Te$_4$ thin film without phosphorus in the temperature range $T=2$–$20$ K~\cite{Gao2024ne}.  [(e) and (f)] For a 6-septuple-layer MnBi$_2$Te$_4$ thin film without phosphorus in the temperature range $T=2$–$10$ K \cite{Yan2023Nature}. [(g) and (h)] For a 4-septuple-layer MnBi$_2$Te$_4$ thin film in the temperature range $T=1.6$–$10$ K \cite{Yan2023Nature}. In (h), the weight of the QMD mechanism is multiplied by 200 times for demonstration and non-QMD is the summation of the weights of the 2SJ, 2SK, SSK, and Drude mechanisms. Different from other panels, the weights in (h) is not displayed in terms of the polynomials of $\sigma_{xx}$ but in terms of the mechanisms.} \label{Fig:Scaling}
\end{figure*}

In Fig.~\ref{Fig:Scaling} we present the fitting results for a series of experiments on MnBi$_2$Te$_4$ thin films \cite{Xu2023Science,Yan2023Nature,Gao2024ne}, which preserve $\mathcal{PT}$ symmetry, therefore QMD is expected to dominate the nonlinear Hall effect.
Figs.~\ref{Fig:Scaling}(a) and (b) are for a 6-septuple-layer MnBi$_2$Te$_4$ thin film strained by black phosphorus \cite{Xu2023Science}. The fitting shows a dominant $\mathcal{C}_0$ term, indicating the dominance of QMD in this experiment according to Table \ref{Tab:scaling}. Figs.~\ref{Fig:Scaling}(c)--(f) are for two devices of 6-septuple-layer MnBi$_2$Te$_4$~\cite{Yan2023Nature,Gao2024ne} in the absence of the black phosphorus. In both devices, the mechanisms are found to be from both QMD and a disorder-induced Drude mechanism, although the two devices are quite different in the magnitudes of the linear longitudinal and second-order Hall conductivities. Figs.~\ref{Fig:Scaling}(g) and (h) are for a 4-septuple-layer MnBi$_2$Te$_4$ thin film \cite{Yan2023Nature}, which indicates that Drude and QMD are the dominant mechanisms, while 2SJ, 2SK, and SSK make only minor contributions.

The substantially different dominant mechanisms in the devices could be understood in terms of their symmetries. 
The unstrained MnBi$_2$Te$_4$ thin film has a three-fold rotational $C_3$ symmetry, which can be broken by strain. The $C_3$ symmetry forbids geometric mechanisms \cite{Du2019NC}, including the BCD and QMD mechanisms. The 6-septuple-layer devices in the absence of black phosphorus likely experience extrinsic $C_3$ symmetry breaking (e.g., substrate-induced strain or
inhomogeneity), allowing QMD to emerge as one of the dominant mechanisms. In summary, the fitting results in Fig.~\ref{Fig:Scaling} support QMD as the dominant mechanism of nonlinear Hall effect in most devices.

We also apply the scaling law to $\mathcal{T}$-even devices to identify the BCD mechanism. Fig.~\ref{Fig:Scaling-T-even} shows the fitting results for two devices of transition metal dichalcogenides, one is WTe$_2$, the other is MoTe$_2$. In both cases, the nonlinear Hall conductivity is found to arise from both BCD and disorder (SJ and SK) mechanisms. More importantly, our scaling analysis can quantitatively distinguish the BCD contribution from the SJ and SK contributions.

\section{Strategy and reliability of fitting}

\begin{figure}[htpb]
\centering
\includegraphics[width=0.45\textwidth]
{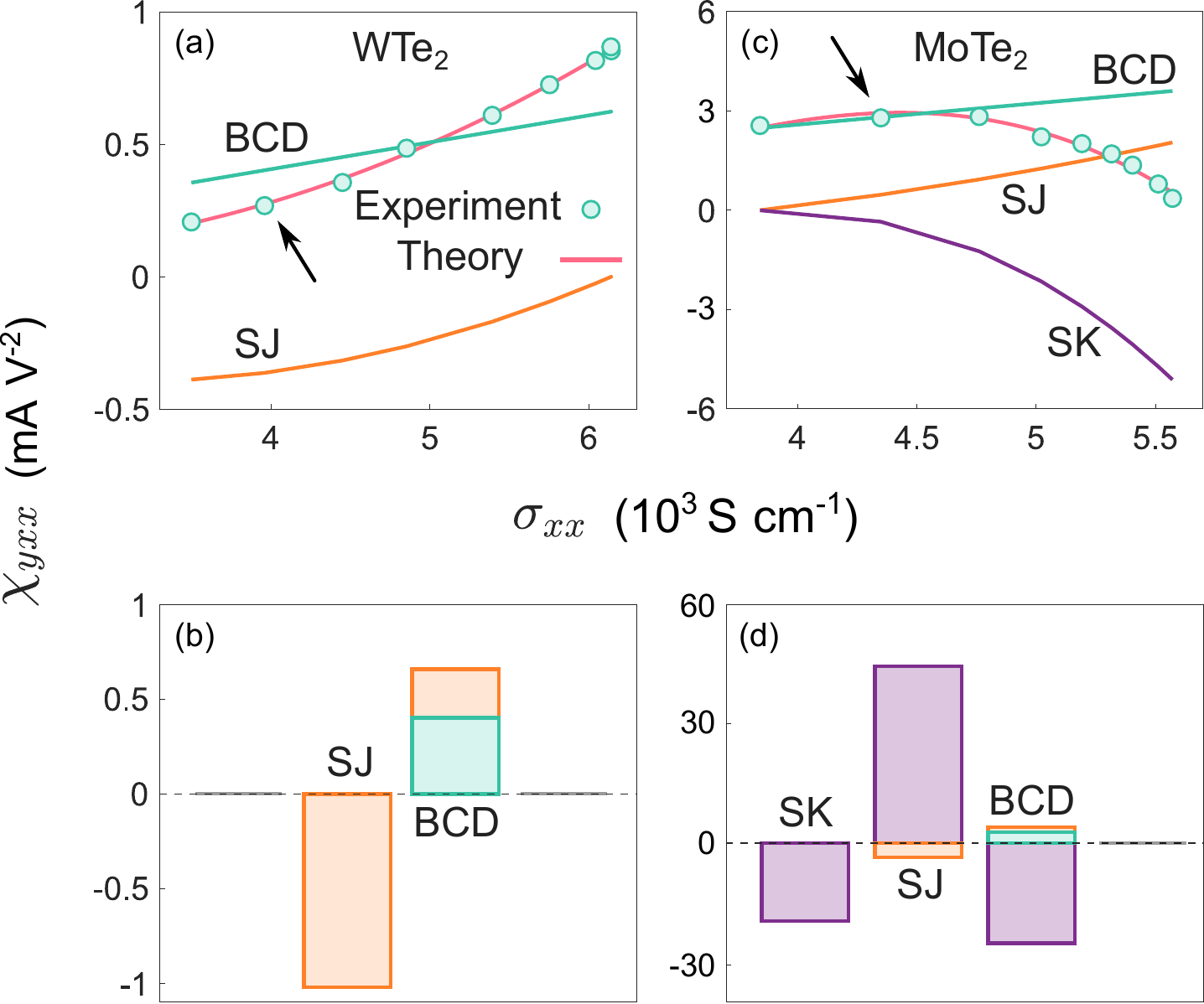}
   \caption{Scaling analysis of the mechanisms for the nonlinear Hall effect in $\mathcal{T}$-even systems. (a) Fitted nonlinear Hall conductivity $\chi_{yxx}$ (red solid curve) as a function of the linear longitudinal conductivity $\sigma_{xx}$ for the experimental data (circles) of a device of WTe$_2$ in the temperature range $T=2$–$100$ K \cite{Kang2019NatMat}.  (b) Weights of the five $\mathcal{C}_i\sigma_{xx}^i$ terms in Eq.~(\ref{eq:scaling-dynamic}) for the data from (a) at $\sigma_{xx}$ = 3.96$\times10^3$ S$\cdot$cm$^{-1}$ (indicated by the arrow). 
[(c) and (d)] For a device of MoTe$_2$ in the temperature range $T=2.5$–$100$ K \cite{ma22nc}.  } \label{Fig:Scaling-T-even}
\end{figure}

To reduce the difficulty of fitting with too many parameters, our strategy is 
to discard statistically insignificant
$\mathcal{C}_n \sigma_{xx}^n$ terms when $|\mathcal{C}_n| < 2\delta_n$, where $\delta_n$ is the standard error, because they are statistically indistinguishable from zero within $\delta_n$. After eliminating the insignificant terms, the scaling law can be rewritten in terms of the mechanisms, to quantify the ratios of the contributing mechanisms. For MnBi$_2$Te$_4$ thin films in Fig.~\ref{Fig:Scaling}, the scaling law becomes
\begin{eqnarray}\label{Eq:scaling-Drude-QMD}
\chi_{yxx}
=\mathcal{C}_\mathrm{Drude}\sigma_{xx}^2+\mathcal{C}_\mathrm{QMD}
\end{eqnarray}
for the 6 septuple layers, and
\begin{eqnarray}
\chi_{yxx}
&=&\mathcal{C}_\mathrm{2SK} (1-\frac{\sigma_{xx}}{\sigma_{xx0}})^4+\mathcal{C}_\mathrm{SSK} (1-\frac{\sigma_{xx}}{\sigma_{xx0}})^3
\notag\\
&&+\mathcal{C}_\mathrm{2SJ}(1-\frac{\sigma_{xx}}{\sigma_{xx0}})^2+\mathcal{C}_\mathrm{Drude}\sigma_{xx}^2\nonumber\\
&&+\mathcal{C}_\mathrm{QMD}    
\end{eqnarray}
for the 4 septuple layers, respectively.  For the 6 septuple layers strained by black phosphorus \cite{Xu2023Science} in Fig.~\ref{Fig:Scaling}(a), $\mathcal{C}_\mathrm{QMD}=13.3\pm0.005$, where the $\pm$ sign gives the standard error. For the 6 septuple layers \cite{Gao2024ne} in Fig.~\ref{Fig:Scaling}(c), $\mathcal{C}_\mathrm{QMD}=(-1.245\pm0.079)\times10^{-1}$ and $\mathcal{C}_\mathrm{Drude}=(1.725\pm0.051)\times10^{-2}$. For the 6 septuple layers \cite{Yan2023Nature} in Fig.~\ref{Fig:Scaling}(e), $\mathcal{C}_\mathrm{QMD}=(-3.74\pm0.317)\times10^{3}$ and $\mathcal{C}_\mathrm{Drude}=(5.59\pm1.12)\times10^{-3}$. For the 4 septuple layers \cite{Yan2023Nature} in Fig.~\ref{Fig:Scaling}(g), $\mathcal{C}_\mathrm{QMD}=(3.08\pm1.77)\times10^{2}$, $\mathcal{C}_\mathrm{Drude}=(-2.38\pm0.765)\times10^{-2}$, $\mathcal{C}_\mathrm{2SJ}=(1.38\pm0.592)\times10^{4}$, $\mathcal{C}_\mathrm{2SK}=(7.24\pm7.04)\times10^{4}$, and $\mathcal{C}_\mathrm{SSK}=(5.36\pm3.51)\times10^{4}$. $\sigma_{xx0}$ is the longitudinal linear conductivity at zero temperature (or the lowest temperature). Similarly, for the $\mathcal{T}$-even cases in Fig.~\ref{Fig:Scaling-T-even}, the scaling law reduces to
\begin{eqnarray}\label{eq:scaling-dynamic-T-even}
\chi_{yxx}
&=&\mathcal{C}_\mathrm{SK} \sigma_{xx}(1-\frac{\sigma_{xx}}{\sigma_{xx0}})^2+\mathcal{C}_\mathrm{SJ} \sigma_{xx}(1-\frac{\sigma_{xx}}{\sigma_{xx0}})\notag\\
&&+\mathcal{C}_\mathrm{BCD} \sigma_{xx}+\mathcal{C}_\mathrm{S},
\end{eqnarray} 
where $\mathcal{C}_\mathrm{S}$ is a fitting parameter for static scattering \cite{Du2019NC}. 
For WTe$_2$ \cite{Kang2019NatMat} in Fig.~\ref{Fig:Scaling-T-even}(a), $\mathcal{C}_\mathrm{BCD}=1.01\pm0.232$,  $\mathcal{C}_\mathrm{SJ}=-2.58\pm0.378$, and $\mathcal{C}_\mathrm{S}=2.35\pm1.45$. For MoTe$_2$ \cite{ma22nc} in Fig.~\ref{Fig:Scaling-T-even}(c),  $\mathcal{C}_\mathrm{BCD}=6.47\pm0.347$,  $\mathcal{C}_\mathrm{SK}=-45.2\pm5.64$, and $\mathcal{C}_\mathrm{SJ}=-8.16\pm2.90$. Besides the standard error, the reliability of our fitting results is also quantified by the coefficient of determination $\mathit{R}^2$ \cite{steel1960principles}, which   $\in[0,1]$ and $\mathit{R}^2=1$ indicates a perfect fit. For the MnBi$_2$Te$_4$ thin films, $\mathit{R}^2=0.99392$ \cite{Gao2024ne} and 0.88319 \cite{Yan2023Nature} for the 6 septuple layers, respectively, and $0.94360$ for the 4 septuple layers \cite{Yan2023Nature}. For the 6 septuple layers strained by black phosphorus, the $\mathcal{C}_0$ term dominates, so $\mathit{R}^2$ is not applicable. For WTe$_2$ and MoTe$_2$, $\mathit{R}^2=0.99836$ and 0.98831, respectively. All the values of $\mathit{R}^2$ are close to 1, indicating the reliability of the fitting results. Moreover, the procedure is validated by the stability of extracted dominant mechanisms across the conductivity range.

\section{Symmetry and Nonzero nonlinear conductivity elements}
The $C_3$ symmetry addressed in Fig.~\ref{Fig:Scaling} can be generalized to all magnetic point groups. The point-group symmetries of $\mathcal{T}$-even mechanisms of the second-order nonlinear transport have been addressed in the previous work \cite{Du2021NC}. Here, we focus on the $\mathcal{T}$-odd mechanisms, which can be classified into three types according to their symmetry properties when exchanging the indices for the input electric fields and output current
\begin{eqnarray}
\mathrm{2SK,2SJ,SSK} &:& \chi_{abc}=\chi_{acb},\nonumber\\
\mathrm{QMD,BSJ,BSK} &:&  \chi_{abc}=-\chi_{cba},\nonumber\\
\mathrm{Drude} &:&\chi_{abc}=\chi_{bca}=\chi_{cab},
\end{eqnarray}
where $\{a,b,c\}\in \{x,y,z\}$. We can use the symmetry properties to determine the nonzero nonlinear conductivity elements $\chi_{abc}$ for all the crystals classified by the magnetic point groups, as shown in Table \ref{Tab:Symmetry} for 2D cases. Compared to the point groups, the magnetic point groups \cite{Litvin13Magnetic} can also describe crystals without $\mathcal{T}$ symmetry.

Note that the QMD mechanism \cite{Su2023NSR,ZhuHY24arXiv} has fewer nonzero elements than the 2SK, 2SJ, and SSK mechanisms, because of the antisymmetric constraint $\chi_{abc}$ = $-\chi_{cba}$. The unstrained \cite{Yan2023Nature} and strained \cite{Xu2023Science} MnBi$_2$Te$_4$ thin films are classified in the $\bar{3}'m'$ and $2/m'$ magnetic point groups, respectively.

\begin{table}[htbp]
\caption[2D nonzero quadratic nonlinear conductivity elements for magnetic point groups]{Symmetry-allowed components of the nonlinear conductivity tensor. Nonzero nonlinear conductivity elements $\chi_{abb}$ on the $x$-$y$ plane (defined as $J_a$ = $\chi_{abb} E_b^2$, $\{a,b\}\in \{x,y\}$) for different mechanisms. In 2D, only 9 out of the 122 magnetic point groups \cite{Litvin13Magnetic} have nonzero elements. In 3D, 20 magnetic point groups have nonzero conductivity elements.}\label{Tab:Symmetry}
\centering
\small\setlength{\tabcolsep}{0.2cm}
\begin{tabular}{ccc}
\hline
\textbf{Groups}  & \textrm{2SK/2SJ/SSK/Drude}   & \textrm{QMD/BSJ/BSK}  \\
\hline
\(\begin{matrix}
\bar{1}^{\prime},~2^\prime/\text{m}
\end{matrix} \)
&
\(\begin{matrix}
\chi_{xxx}		&	\chi_{xyy}		\\
\chi_{yxx}		&	\chi_{yyy}		
\end{matrix} \)
&
\(\begin{matrix}
\chi_{yxx}		&	\chi_{xyy}		
\end{matrix} \)
\\
\hline
\rule{0pt}{2.5ex}
\(\begin{matrix}
2/\text{m}^\prime,~\text{m}^\prime\text{m}\text{m}
\end{matrix} \)
&
\(\begin{matrix}
\chi_{yxx}		&	\chi_{yyy}		
\end{matrix} \)
&
\(\begin{matrix}
\chi_{yxx}	
\end{matrix} \)
\\
\hline
\rule{0pt}{4ex}
\(\begin{matrix}
\bar{3}^\prime\text{m}, \bar{3}^\prime\text{m}^\prime,\\
6^\prime/\text{m}\text{m}\text{m}^\prime
\end{matrix} \)
&
\(\begin{matrix}
\chi_{yxx}=-\chi_{yyy}		
\end{matrix}\)
&
$\times$ 
\\
\hline
\rule{0pt}{2.5ex}
\(\begin{matrix}
\bar{3}^\prime,~6^\prime/\text{m}
\end{matrix} \)
&
\(\begin{matrix}
\chi_{xyy}=-\chi_{xxx} 		\\
\chi_{yxx}=-\chi_{yyy}		
\end{matrix}\)
&
$\times$
\\
\hline
\end{tabular}
\end{table}

\section{Experimental implications}
Our work  suggests several potential future directions. The scaling law in Eq.~(\ref{eq:scaling-dynamic}) and Table \ref{Tab:scaling} can be applied to identify mechanisms in future experiments on nonlinear transport. The nonzero elements of nonlinear conductivity in Table \ref{Tab:Symmetry} can guide future experiments in more materials. The theory can be applied to quantitatively study nonlinear transport in more systems and application scenarios, such as rectification, piezoelectric, and memory devices.

\section{Derivation of nonlinear conductivity formulas}


We briefly describe the calculation of the nonlinear conductivity formulas. We start with the uniform Boltzmann equation \cite{Du2019NC,Sinitsyn2008JPCM}
\begin{eqnarray}
\partial f_{l}/\partial t+\dot{\mathbf{k}} \cdot  \partial f_{l}/\partial \mathbf{k}=\mathcal{I}_{e l}\{f_{l}\}
\end{eqnarray}
of the distribution function $f_l$, where $\mathcal{I} _{el}\{ f_{l} \} $ accounts for the elastic disorder scattering, $l$ stands for the band index $\gamma$ and wave vector $\mathbf{k}$, $\dot{\mathbf{k}}=-(e/\hbar) \mathbf{E}$, and $\mathbf{E}$ is the electric field. By decoupling the distribution function and scattering according to different mechanisms ($in$ for disorder-irrelevant, $sj$ and $sk$ for side jump and skew scattering, respectively)
\begin{eqnarray}\label{eq:distribution}
\centering
 f_{l}&=&f_{l}^{i n}+\delta f_{l}^{s j}+\delta f_{l}^{s k}+\delta f_{l}^{2s j}\nonumber\\
 &&+\delta f_{l}^{2s k}+\delta f_{l}^{s k,sj},\nonumber\\
\mathcal{I} _{el}\{ f_{l} \} &=&\mathcal{I}^{in} +\mathcal{I}^{sj}  +\mathcal{I}^{sk}  +\mathcal{I}^{sk,sj}, 
\end{eqnarray}
$f_l$ can be found up to the second order of $\mathbf{E}$, where $\delta f_{l}^{2s j}+\delta f_{l}^{2s k}+\delta f_{l}^{s k,sj}$ and $\mathcal{I}^{sk,sj}$ are crucial for new findings. The nonlinear conductivity $\chi_{abc}$ can be extracted from the electric current density
$\mathbf{J}(\mathbf{E})=-e \sum_{l} \dot{\mathbf{r}}_{l} f_{l}$,
where the velocity
\begin{eqnarray}
\dot{\mathbf{r}}_{l}=\mathbf{v}_{l}-\dot{\mathbf{k}} \times\left( {\Omega}_{l}+ \nabla_\mathbf{k}\times\boldsymbol{\mathcal{G}}\mathbf{E}\right)+\mathbf{v}_{l}^{sj}
\end{eqnarray}
includes the terms from the group velocity $\mathbf{v}_l$, Berry curvature ${\Omega}$ \cite{Xiao2010RMP}, Berry connection polarizability  $\boldsymbol{\mathcal{G}}$, and SJ velocity $\mathbf{v}^{sj}$.

\begin{table}[htbp]
\centering
\caption{Nonlinear conductivity formulas at the
quantum metric order. Formulas of the nonlinear conductivity at quantum-metric order, including QMD, Drude, BSK, BSJ, 2SJ, 2SK, and SSK. The hierarchy and physical pictures of them are illustrated in Fig.~\ref{Fig:Mechanisms}. The quantities and symbols are defined below Eq.~(\ref{Eq:Total}). The conductivity formulas are also given in Ref.~\cite{Huang25prb}.} \label{Tab:formulas}
\small\begin{tabular}{l}
\hline
 {$\chi^\mathrm{QMD}=e^2\sum_l{\left( v_{l}^{a}\mathcal{G} _{l}^{cb}-v_{l}^{c}\mathcal{G} _{l}^{ab} \right)} \partial_{\epsilon_{l}} f^{(0)}_l$} \cite{GaoY14prl,Xiao2021PRL,Yang2021PRL,Xiao25prl,Qiang25AS},\\
 $\chi^\mathrm{Drude}=-\frac{\tau^2e^3 }{2\hbar^2} \sum_l v_l^a\partial_{\mathbf{k}}^b\partial_{\mathbf{k}}^cf_l^{(0)}$  \cite {Fu2015PRL},\\
 $\chi^\mathrm{BSK}_1=\frac{\tau ^2e^3}{2\hbar ^2}\sum_{l l^{\prime}}{\varepsilon ^{acd}\Omega_{l^{\prime}}^{d}}\varpi_{ll^{\prime}}^{(3A)}\partial _{\mathbf{k}}^{b}f_{l}^{(0)}$ \cite{Song2023PRL},\\
 $\chi_2^{\mathrm{BSK}}=\frac{\tau ^2e^3}{2\hbar ^2}\sum_{l l^{\prime}}{\varepsilon ^{acd}\Omega_{l^{\prime}}^{d}}\varpi_{ll^{\prime}}^{(4A)}\partial _{\mathbf{k}}^{b}f_{l}^{(0)}$,\\
 $\chi^{\mathrm{BSJ}}=\frac{\tau e^3}{2\hbar}\sum_l{\varepsilon ^{acd}\Omega _{l}^{d}v_{l}^{sj,b}  \partial_{\epsilon_{l}} f_l^{(0)}}$,\\
 $\chi_1^{2 \mathrm{SK}}=\frac{e^3 \tau^4}{2 \hbar^2} \sum_{l l^{\prime} l^{\prime \prime}}\left(\partial_{\mathbf{k}^{\prime \prime}}^c v_{l^{\prime \prime}}^a\right) \varpi_{l^{\prime} l}^A \varpi_{l^{\prime \prime} l^{\prime}}^A \partial_{\mathbf{k}}^b f_l^{(0)}$,\\
 $\chi_2^{2 \mathrm{SK}}=\frac{e^3 \tau^4}{2 \hbar^2} \sum_{l l^{\prime} l^{\prime \prime}}v_{l^{\prime \prime}}^a\left(\partial_{\mathbf{k}}^c-\partial_{\mathbf{k}^{\prime}}^c\right) \varpi_{l^{\prime} l}^A \varpi_{l^{\prime \prime} l^{\prime}}^A \partial_{\mathbf{k}}^b f_l^{(0)}$,\\
  $\chi_1^{2 \mathrm{SJ}}=\frac{\tau^2 e^3}{2} \sum_{l l^{\prime}}\left(v_{l^{\prime}}^a+v_l^a\right) \bar{O}_{l^{\prime} l}^c v_{l}^{sj,b} \partial_{\epsilon_{l}} f_l^{(0)}$,\\
  $\chi_2^{2 \mathrm{SJ}}=\frac{\tau ^2e^3}{2\hbar}\sum_{ll^{\prime}}{( v_{l}^{sj,a}+v_{l^{\prime}}^{sj,a}) \bar{O}_{ll^{\prime}}^{c}\partial _{\mathbf{k}}^{b}f_{l}^{(0)}}$,\\
  $\chi_3^{2 \mathrm{SJ}}=-\frac{\tau ^2e^3}{2\hbar}\sum_l{\partial_{\mathbf{k}}^{b}v_{l}^{sj,a}}v_{l}^{sj,c}
\partial_{\epsilon_{l}}f_{l}^{(0)}$,\\
 $\chi_1^{ \mathrm{SSK}}=\frac{\tau^2 e^3}{2 \hbar} \sum_{l l^{\prime}}\left(v_l^a+v_{l^{\prime}}^a\right) \tilde{O}_{l^{\prime} l}^c \partial_{\mathbf{k}}^b f_l^{(0)}$,\\
 $\chi_2^{ \mathrm{SSK}}=\frac{\tau^3 e^3}{2 \hbar} \sum_{l l^{\prime} l^{\prime \prime}}\left(v_{l^{\prime \prime}}^a+v_{l^{\prime}}^a\right) \varpi_{l l^{\prime}}^A \bar{O}_{l^{\prime \prime} l^{\prime}}^c \partial_{\mathbf{k}}^b f_l^{(0)}$,\\
 $\chi_3^{ \mathrm{SSK}}=\frac{\tau^3 e^3}{2 \hbar} \sum_{l l^{\prime} l^{\prime \prime}}\left(\varpi_{l^{\prime} l}^A+\varpi_{l^{\prime} l^{\prime \prime}}^A\right) v_{l^{\prime}}^a \bar{O}_{l l^{\prime}}^c \partial_{\mathbf{k}}^b f_l^{(0)}$\\
 $\quad\quad\quad+ \frac{\tau^3 e^3}{2 \hbar}\sum_{l l^{\prime}} v_{l^{\prime}}^a v_l^{s j, b}\left(\partial_{\mathbf{k}}^c-\partial_{\mathbf{k}^{\prime}}^c\right) \varpi_{l^{\prime} l}^A \partial_{\epsilon_l} f_l^{(0)}$,\\
 $\chi_4^{\mathrm{SSK}}=\frac{\tau^3 e^3}{2 \hbar^2} \sum_{ll^{\prime}} v_{l^{\prime}}^{s j, a}\left(\partial_{\mathbf{k}}^c-\partial_{\mathbf{k}^{\prime}}^c\right) \varpi_{l^{\prime} l}^A \partial_{\mathbf{k}}^b f_l^{(0)}$. \\
\hline
\end{tabular}
\end{table}

The nonlinear conductivity formula at quantum-metric order in Eq.~(\ref{Eq:chi_abc}) can be decomposed into
\begin{eqnarray}\label{Eq:Total}
\chi_{abc}&=&\sum_{i=1}^2\chi_i^{\mathrm{2SK}}+\sum_{i=1}^3\chi_i^{\mathrm{2SJ}}
+\sum_{i=1}^4\chi_i^{\mathrm{SSK}}
+\chi^{\mathrm{BSJ}}
\nonumber\\
&&
+\sum_{i=1}^2\chi_i^{\mathrm{BSK}}
+\chi^\mathrm{Drude}
+\chi^{\mathrm{QMD}}.
\end{eqnarray}
Their explicit forms are given in Table \ref{Tab:formulas}, where $-e$ is the electron charge, the scattering time $\tau$ is defined as
\begin{eqnarray}\frac{1}{\tau}
\approx \frac{1}{\tau_{\bf k}} \equiv  \sum_{\bf k'}\varpi_{\mathbf{k}^{\prime} \mathbf{k}}^{(2)}\left[1-\cos \left(\phi-\phi^{\prime}\right)\right] \delta\left(\epsilon_{\mathrm{F}}-\epsilon_{l}\right),
\end{eqnarray}
$\varepsilon^{abc}$ is the antisymmetric tensor, the Fermi distribution
$f_{l}^{(0)}=\{ \exp[(\epsilon_{l}-\epsilon_{F})/k_B T]+1\}^{-1}$,
with the Fermi energy $\epsilon_{F}$, $l$ stands for $(\gamma,\mathbf{k})$, with the band index $\gamma$ and wave vector 
$\mathbf{k}=|k|\left( \cos \phi ,\sin \phi \right)
$ in 2D,
the Berry curvature \cite{Xiao2010RMP}
\begin{eqnarray}
\Omega^{a}_{l}\equiv-2\varepsilon^{abc}\sum_{\gamma^{\prime}\neq \gamma}\frac{\mathrm{Im}\langle \gamma|\partial^{b}_\mathbf{k}\hat{\mathcal{H}}|\gamma^\prime\rangle\langle \gamma^\prime|\partial^{c}_\mathbf{k}\hat{\mathcal{H}}|\gamma\rangle}{(\epsilon^\gamma_{\mathbf{k}}-\epsilon^{\gamma^{\prime}}_{\mathbf{k}})^{2}},
\end{eqnarray}
$\partial^{b}_\mathbf{k}$$\equiv$$\partial/\partial k_{b}$, $\hat{\mathcal{H}}$ is the Hamiltonian, $\epsilon_{l}$ is the eigenenergy, group velocity $v_{l}^a$ = $(1/\hbar) \partial \epsilon_{l}/\partial k_a$,
$\partial_{\epsilon_{l}}\equiv \partial /\partial \epsilon_{l}$, $\partial_{\epsilon_{l}} f^{(0)}_l $ means summing those states near the Fermi surface. The SJ velocity
$v_{l}^{sj,a}$ = $\sum_{l'}\varpi^{S}_{ll'}\delta r^{a}_{l'l}$,
$\varpi^{S}_{ll'}$ ($\varpi _{l'l}^{A}$) is the symmetric (antisymmetric) scattering rate between states $l$ and $l'$, the coordinates shift
\begin{eqnarray}
\delta r^a_{ll'}\equiv i\langle l |\partial^a_\mathbf{k} |l\rangle-i\langle l'|\partial^{a}_\mathbf{k'}|l'\rangle-(\partial^{a}_\mathbf{k}+\partial^{a}_\mathbf{k'})\arg(V_{ll'}),
\end{eqnarray}
 with $V_{ll'}$=$\langle l|\hat{V}_{imp}|l'\rangle$ \cite{Du2019NC,Sinitsyn2008JPCM}, where the $\delta$-correlated random potential
 $\hat{V}_{imp}(\mathbf{r})=\sum_{i}V_{i}\delta(\mathbf{r}-\mathbf{R}_{i})$
 can give the Gaussian $\langle V^{2}_{i}\rangle_{dis}=V^{2}_{0}$,
and non-Gaussian correlations $\langle V^{3}_{i}\rangle_{dis}$=$V^{3}_{1}$ \cite{Du2019NC,Du2021NC,Sinitsyn2008JPCM}, leading to the intrinsic SK $\varpi_{l'l}^{(4A)}$ and extrinsic SK $\varpi_{l'l}^{(3A)}$ parts to 
\begin{eqnarray}\label{Eq:skew-scattering}
\varpi _{l'l}^{A}=\varpi _{l'l}^{3A}+\varpi _{l'l}^{4A}.    
\end{eqnarray}
$\bar{O}_{ll'}^{a}$ ($\tilde{O}_{ll'}^{a}$) is the symmetric (antisymmetric) part of  \cite{Du2019NC}
\begin{eqnarray}\label{Eq:O}
O^{a}_{ll'}\equiv(2\pi/\hbar)|T_{ll'}|^{2}\delta r^{a}_{ll'}\partial_{\epsilon_{l}}\delta(\epsilon_{l}-\epsilon_{l'}),
\end{eqnarray}
and $T_{ll'}$ is the T-matrix \cite{Mahan2013}. Alternatively, the calculation can be based on the density matrix \cite{Atencia22prr,Atencia23prbl,Mehraeen24prb}.

\section{ Nonlinear conductivity of QMD}
In the nonlinear conductivity formula of the QMD mechanism $\chi^\mathrm{QMD}$ in Table \ref{Tab:formulas},  the Berry connection polarizability tensor reads \cite{GaoY14prl,Xiao2021PRL,Yang2021PRL} 
\begin{eqnarray}\label{Eq:berry-connection-polarizability}
\mathcal{G}^{a b}_{\gamma,\mathbf{k}}\equiv 2 e \mathrm{Re} \sum_{\gamma^{\prime} \neq \gamma}  A^{\gamma \gamma^{\prime}}_{a,\mathbf{k}} A^{\gamma^{\prime} \gamma}_{b,\mathbf{k}}/(\epsilon^\gamma_{\mathbf{k}}-\epsilon^{\gamma^{\prime}}_{\mathbf{k}})
\end{eqnarray}
is related to the quantum metric tensor \cite{Provost80cmp,Resta11epjb}
\begin{eqnarray}\label{Eq:QMD-expression} 
g_{\gamma,\mathbf{k}}^{ab} \equiv \mathrm{Re} \sum_{\gamma\neq \gamma'} A^{\gamma \gamma^{\prime}}_{a,\mathbf{k}} A^{\gamma^{\prime} \gamma}_{b,\mathbf{k}},
\end{eqnarray}
where the Berry connection $A^{\gamma \gamma'}_{a,\mathbf{k}}\equiv i\langle \gamma |\partial_\mathbf{k}^a | \gamma'\rangle $ \cite{Xiao2010RMP}. For two-band systems, they are explicitly related as
\begin{eqnarray}
\mathcal{G}^{a b}_{\gamma,\mathbf{k}}=2e  g_{\gamma,\mathbf{k}}^{ab}/(\epsilon^\gamma_{\mathbf{k}}-\epsilon^{\gamma^{\prime}}_{\mathbf{k}}).    
\end{eqnarray}

\section{Symmetry of nonlinear conductivity formulas}
In Eq.~(\ref{Eq:berry-connection-polarizability}), the velocity $v$ is odd under both $\mathcal{T}$ and $\mathcal{P}$ operations, while the Berry curvature polarizability $\mathcal{G}$ is even under both $\mathcal{T}$ and $\mathcal{P}$ operations, so  $v\mathcal{G}$ is odd under both $\mathcal{T}$ and $\mathcal{P}$ operations but even under the joint
$\mathcal{PT}$ symmetry. Therefore, to have a nonzero $\chi^\mathrm{QMD}$, both the $\mathcal{T}$ and $\mathcal{P}$ must be broken, while $\chi^\mathrm{QMD}$ can survive the $\mathcal{PT}$ symmetry. We can use Table \ref{Tab:Symmetry-quantity} to check that all formulas in Table \ref{Tab:formulas} have the same symmetries of  $\chi^\mathrm{QMD}$.

\begin{table}[htbp]
\caption{Symmetry properties of quantities in non-
linear conductivity formulas. Whether the quantities in the nonlinear conductivity formulas in Table \ref{Tab:formulas} change sign {($-$) or not ($+$)}, under $\mathcal{P}$, $\mathcal{T}$, and the joint $\mathcal{PT}$ symmetry operations.}\label{Tab:Symmetry-quantity}
\setlength{\tabcolsep}{0.15cm}
\centering
\begin{tabular}{cccccccccc}
\hline
   & $\partial_a$ &  $v^a_l$ &  $\Omega^a_l$  &  $\delta r^{a}_{ll'}$  &   $v_{a,l}^{sj}$  & $\bar{O}^{a}_{ll'}$  & $\tilde{O}^{a}_{ll'}$ &  $\varpi _{ll'}^{S}$ &  $\varpi _{ll'}^{A}$  \\
  \hline
$\mathcal{P}$                      &  $-$ &  $-$ & $+$ &  $-$ & $-$ & $-$ & $-$  & $+$ & $+$ \\
$\mathcal{T}$                      & $-$  & $-$  & $-$ &  $+$ & $+$ & $+$ & $-$  & $+$ & $-$ \\
$\mathcal{PT}$                      &  $+$ &  $+$ & $-$ &  $-$ & $-$ & $-$ & $+$  & $+$ & $-$ \\
    \hline
\end{tabular}
\end{table}

Take the formula of $\chi_4^{\mathrm{SSK}}$ in Table \ref{Tab:formulas} for example, in which 
$v_{l^{\prime}}^{s j, a}\left(\partial_{\mathbf{k}}^c-\partial_{\mathbf{k}^{\prime}}^c\right) \varpi_{l^{\prime} l}^A \partial_{\mathbf{k}}^b$ gives $(-1)$ $(-1)$ $(+1)$ $(-1)$ = $-1$ upon $\mathcal{P}$, $(+1)$ $(-1)$ $(-1)$ $(-1)$ $=$ $-1$ upon $\mathcal{T}$, and $(-1)$ $(+1)$ $(-1)$ $(+1)$ $=$ $1$ under the joint $\mathcal{PT}$ operation, so it has the same symmetry of $\chi^{\mathrm{QMD}}$. The symmetry of the other formulas can be shown similarly.

\section{Conflict of interest}
The authors declare that they have no conflict of interest.

\section{Acknowledgements}
We thank Xiangang Wan and Xiaoqun Wang for helpful discussions. This work was supported by the National Key R$\&$D Program of China (2022YFA1403700), Quantum Science and Technology–National Science and Technology Major Project (2021ZD0302400), the National Natural Science Foundation of China (12525401, 12350402, 12374041), Guangdong Basic and Applied Basic Research Foundation (2023B0303000011), Guangdong Provincial Quantum Science Strategic Initiative (GDZX2201001 and GDZX2401001), the Science, Technology and Innovation Commission of Shenzhen Municipality (ZDSYS20190902092905285), High-level Special Funds (G03050K004), and Center for Computational Science and Engineering of SUSTech.

\section{Author contributions}
Zhen-Hao Gong did the calculations with assistance from Z. Z. Du, Hai-Peng Sun, and Hai-Zhou Lu; Zhen-Hao Gong, Z. Z. Du, and Hai-Zhou Lu wrote the manuscript with assistance from Hai-Peng Sun and X. C. Xie; Hai-Zhou Lu and X. C. Xie supervised the project.

\bibliographystyle{apsrev4-2-title}
\bibliography{reference}

\end{document}